\documentclass[12pt,amsbsy,epsf]{article}
\usepackage{graphicx}
\usepackage{amssymb}

\addtocounter{page}{0} \tolerance=5000

\newcommand{\beq}{\begin{equation}}
\newcommand{\eeq}{\end{equation}}
\newcommand{\beqn}{\begin{eqnarray}}
\newcommand{\eeqn}{\end{eqnarray}}

\newcommand{\n}{\mathbf{n}}
\newcommand{\p}{\mathbf{p}}
\newcommand{\pp}{\mathbf{p}'}

\newcommand{\bj}{\mathbf{j}}
\newcommand{\bk}{\mathbf{k}}

\newcommand{\bq}{\mathbf{q}}
\newcommand{\br}{\mathbf{r}}
\newcommand{\bS}{\mathbf{S}}

\newcommand{\vA}{\mathbf{A}}
\newcommand{\va}{\mathbf{a}}
\newcommand{\vI}{\mathbf{I}}

\newcommand{\Na}{\mbox{{\boldmath$\nabla$}}}
\newcommand{\si}{\mbox{{\boldmath$\sigma$}}}

\newcommand{\al}{\mbox{${\alpha}$}}
\newcommand{\ep}{\mbox{${\varepsilon}$}}

\begin{document}
\begin{titlepage}
\begin{center}
{\Large \bf  Variations on the deuteron}
\end{center}

\vspace{1cm}

\begin{center}
I.B. Khriplovich\footnote{khriplovich@inp.nsk.su}\\Budker
Institute of Nuclear Physics\\ 630090 Novosibirsk, Russia,\\ and
Novosibirsk University
\end{center}

\vspace{1cm}

\begin{flushright}
{\it To the memory of Arkady Benediktovich Migdal,\\ the teacher
of my teachers}
\end{flushright}

\vspace{1cm}

\begin{abstract}
We consider few problems which are related to the deuteron and
have simple analytical solution. Relation is obtained between the
deuteron electric quadrupole moment and the $np$ scattering
amplitude. The degree of circular polarization of photons is
found for the radiative capture of longitudinally polarized
thermal neutrons. The anapole, electric dipole and magnetic
quadrupole moments of the deuteron are calculated.
\end{abstract}

\vspace{4cm}

\end{titlepage}

\section{Introduction}
Let us recall the classical question by Fermi: ``What plays the
part of a hydrogen atom in this problem?'' As to nuclear physics,
certainly, its hydrogen atom is the deuteron.

It is surprising how many nontrivial problems related to the
deuteron can be solved by means of sufficiently simple, sometimes
truly back-of-the-envelope analytical calculations. The reason is
essentially that the deuteron binding energy, $\ep=2.23$ MeV, is
anomalously small on the nuclear scale. As a result of it, the
deuteron wave function decreases very slowly beyond the range of
nuclear forces. In the exponential asymptotics  $e^{-\kappa r}$ at
long distances the parameter $\kappa$ is small: $\kappa
=\sqrt{m_p\ep}=45.7$ MeV ($m_p$ is the proton mass).
Correspondingly, the typical length at which the wave function
falls down, is large: $\kappa^{-1} = 4.3$ fm, it is much larger
than the range of nuclear forces $r_0 \sim 1$ fm. This allows one
to use for the deuteron wave function the so-called zero-range (of
nuclear forces) approximation where this wave function is
approximated by its asymptotic expression
\begin{equation}\label{d}
\psi_{d}=\sqrt{\frac{\kappa}{2\pi}}\frac{e^{-\kappa r}}{r}.
\end{equation}
Due to the coefficient $\sqrt{\kappa/2\pi}$, this expression is
normalized correctly: $\int d\br \psi_{d}^2=1$.

In the present review some results are considered which were
obtained relatively recently basing on the zero-range
approximation. Being related to the deuteron, these problems
belong nevertheless to somewhat different fields: to the
traditional nuclear physics, and to the problem of nonconservation
of parity and time-reversal invariance.

It is convenient however to discuss at first briefly few classical
problems belonging already to textbooks or simply to physical
folklore. The deuteron ground state contains in line with $^3S_1$
a small admixture of $^3D_1$, which can be in many cases
neglected. The wave function of the triplet $^3S_1$ state of the
continuous spectrum of the system neutron-proton can be written in
the low-energy limit as
\begin{equation}\label{St0}
\psi_{St}=1- \,\frac{\alpha_t}{r},
\end{equation}
where $\alpha_t=5.42$ fm is the triplet scattering length. The
analogous expression for the singlet $^1S_0$ function of the
continuous spectrum is
\begin{equation}\label{Ss0}
\psi_{Ss}=1- \,\frac{\alpha_s}{r}.
\end{equation}
The singlet scattering length is negative and very large:
$\alpha_s=-23.7$ fm. The subscript $S$ at the wave function
denotes here and below an $S$-wave, the subscripts $t$ and $s$
denote triplet and singlet states, respectively. Somewhat more
precise expressions for the wave functions of $S$-states of
continuous spectrum, taking into account the relative momentum $p$
of scattering particles (it is small as compared to $1/r_0$, but
not as compared to $1/\al_{t,s}$), can be conveniently written
as~\cite{mmr}
\begin{equation}\label{St,Ss}
\psi_{St,Ss}={\sin pr \over pr}
-\,\frac{\alpha_{t,s}}{1+i\,p\,\alpha_{t,s}}\,\frac{e^{ipr}}{r}\,.
\end{equation}

The simplest of the mentioned classical results is the relation
between the deuteron binding energy and the triplet scattering
length. From the orthogonality of the triplet wave functions
(\ref{d}) and (\ref{St0}) which correspond to different energies,
one obtains immediately
\beq\label{ak}
\al_t=\,{1 \over \kappa},\quad \mbox{or}\quad \al_t=\,{1
\over\sqrt{m_p\ep}}.
\eeq
It is valid within $\sim 20\%$.

Let us consider now the cross-section of the deuteron
photodisintegration $\gamma d \to n p$. It is natural to start
with the contribution of $E1$ transition~\cite{bp,blp}. The dipole
matrix element is
\beq\label{AE1}
\langle ^3P|\,{\bf r}|^3S \rangle =\,-\,4\,\sqrt{{2\pi\kappa \over
1-\kappa r_t}}\;{{\bf p} \over (\kappa^2+p^2)^2}.
\eeq
The corresponding contribution to the cross-section is
\beq\label{SE1}
\sigma_{E1}=\,{8\pi \alpha \over 3}\,{\kappa  p^3 \over (1-\kappa
r_t)\, (\kappa^2+p^2)^3}.
\eeq
The result of a straightforward calculation with wave function
(\ref{d}) is augmented here with the correction factor  $(1-\kappa
r_t)^{-1/2}$ in matrix element (\ref{AE1}), and correspondingly,
by $(1-\kappa r_t)^{-1}$ in cross-section (\ref{SE1}); here
$r_t=1.76$ fm is the so-called effective interaction radius. The
correction is not so small, $1-\kappa r_t=0.59$. Its origin is as
follows. Matrix element (\ref{AE1}) is dominated by large
distances, and at these distances the account for the effective
radius modifies the asymptotics of the deuteron wave function just
in this way~\cite{blp,ff}. As usual, the cross-section of a $E1$
transition is proportional at the threshold to $p^3$.

Meanwhile, the contribution of $M1$ transition into the deuteron
photodisintegration cross-section decreases in the threshold
region as $p^1$ only, and therefore dominates very close to the
threshold~\cite{blp,fe}. Let us consider this contribution. Since
the orbital angular momentum of the nucleons in the deuteron
equals to zero (we neglect here the admixture of $D$-wave), the
magnetic moment operator reduces here to a purely spin one:
\[ {\bf M}=\,{e \over 2m_p}\,(\mu_p \si_p + \mu_n \si_n). \]
Here $\si_p$ and $\si_n$ are the spin operators of the proton and
neutron, respectively, and $\mu_p=2.79$ and $\mu_n=-1.91$ are
their magnetic moments. Due to the same orthogonality of the wave
functions of the deuteron and $^3S_1$ state of positive energy,
the $M1$ transition goes into the singlet $^1S_0$ state of the
continuous spectrum. A simple calculation~\cite{blp,fe} results in
\beq\label{AM1}
\langle ^1S_0|(\mu_p \sigma_p +\mu_n \sigma_n)_z|^3S_1 \rangle =
{2\sqrt{2\pi\kappa}\,(\mu_p-\mu_n)\,(1-\kappa\al_s) \over
(\kappa^2+p^2)\,(1-ip\,\al_s)}\,,
\eeq
and correspondingly,
\beq\label{SM1}
\sigma_{M1}=\,{2\pi \alpha \over 3}\,{\kappa p\,
(\mu_{p}-\mu_{n})^{2} (1-\kappa\alpha_{s})^{2} \over
m_{p}^{2}\,(\kappa^2+p^2)\,(1+p^2\alpha_{s}^2)}.
\eeq
Though this contribution to the cross-section is enhanced by large
numerical factors, $\mu_{p}-\mu_{n}=4.7$ and
$1-\kappa\alpha_{s}=6.5$, it dominates only within 0.2 MeV above
the threshold.

After this, rather lengthy, Introduction we go over to the main
part of the review.

\section{The $np$-scattering amplitude\\ and the deuteron quadrupole moment}
Just as the triplet scattering length $\al_t$ is related to the
asymptotic behaviour of the deuteron $^3S_1$ wave function at $r
\to \infty$, i.e., to the parameter $\kappa$ (see (\ref{ak})), the
limiting threshold values of the spin-dependent invariant
amplitudes is related to the asymptotic value of the $^3D_1$
admixture in the deuteron wave function~\cite{bkk}.

We start with the standard expression for the $np$ scattering
amplitude, valid under the assumptions of $P$ and $T$ invariance,
as well as of the charge independence of nuclear forces
(see~\cite{wa}):
\beq\label{amp}
f(\pp, \p)=a+b(\si_1\n)(\si_2\n)+c(\si_1+\si_2)\n +
(g+h)(\si_1\n_+)(\si_2\n_+) +(g-h)(\si_1\n_-)(\si_2\n_-).
\eeq
Here $\p$ ¨ $\pp$ are the nucleon relative momenta in the
center-of-mass system, initial and final, respectively; $\si_1$
and $\si_2$ are their spin operators;
\[ \n_{\pm}=\,{\p\pm\pp \over |\p\pm\pp|}\,, \quad
\n={\p\times\pp \over |\p\times\pp|}. \] We get rid of the
structure $(\si_1\n)(\si_2\n)$ in (\ref{amp}) by means of the
obvious relationship $(\si_1\n)(\si_2\n)=\si_1\si_2
-(\si_1\n_+)(\si_2\n_+)-(\si_1\n_-)(\si_2\n_-)$. Then, we use the
momenta $\p$ and $\pp$ instead of the unit vectors $\n_{\pm}$. And
at last, we go over from the operators $\si_1$ and $\si_2$ to the
total spin operator $\bS=1/2(\si_1+\si_2)$ since we are interested
in the triplet scattering. As a result, the triplet $np$
scattering amplitude becomes
\beq\label{amp1}
f(\pp, \p)=-\al_t+\,{1 \over m_p^2}\left\{c'\bS[\pp\times
\p]+g_1\left(\bS(\pp+\p)\right)^2 +
g_2\left(\bS(\pp-\p)\right)^2\right\},
\eeq
Here the triplet scattering length is related to the parameters of
formula (\ref{amp}) as follows: $\al_t=-(a+b)$; the constants $$
c'=\,{2c m_p^2 \over p^2\sin \theta}\,, \quad g_1=\,{(g-b+h)\,
m_p^2 \over 2 p^2 \cos^2(\theta/2)}\,, \quad g_2=\,{(g-b-h)\,
m_p^2 \over 2 p^2 \sin^2(\theta/2)}\,, $$ as well as $\al_t$, are
independent of the scattering angle $\theta$. Expression
(\ref{amp1}) is in fact an expansion of the scattering amplitude
in momenta, where one retains in line with $\al_t$ only those
higher order terms which depend on the spin $S$.

We will go over now to the construction of the deuteron wave
function at large $r$, which includes the $D$ wave admixture. To
this end, we find at first the effective $\delta$-function
potential (pseudopotential) which reproduces in the Born
approximation the scattering amplitude
(\ref{amp1})(see~\cite{ll}). With the correspondence
\[  p_i'p_j' \to -\nabla_i\nabla_j\delta(\br)\,,
p_i'p_j \to -\nabla_i\delta(\br)\nabla_j\,, p_i p_j \to
-\delta(\br)\nabla_i\nabla_j\,, \] we obtain
\beq
U(\br)=\,{4\pi \over m_p}\, \{\al_t\,\delta(\br)+\,{c' \over
m_p^2}\,\ep_{ijk}S_i\,\nabla_j\,\delta(\br)\,\nabla_k +\,{1 \over
2 m_p^2}S_{ij}\,[(g_1+g_2)\,(\nabla_i \nabla_j\delta(\br)+
\delta(\br)\nabla_i \nabla_j)
\eeq
\[  + (g_1-g_2)\,(\nabla_i
\delta(\br)\nabla_j+\nabla_j\delta(\br)\nabla_i)] \}, \] where
\[S_{ij}= S_iS_j+S_jS_i-\,{4 \over
3}\,\delta_{ij}. \]

In the wave function of the scattering problem the outgoing
spherical wave can be presented as
\beq\label{sca}
-\,{m_p \over 4\pi}\,\int d\br'\,{e^{ip|\br-\br'|} \over
|\br-\br'|}\,U(\br')\Psi_0(\br')\,.
\eeq
We assume that the nonperturbed solution $\Psi_0$ is an $S$ wave
one. Then expression (\ref{sca}) reduces to
\beq\label{wf}
\left(\al_t+{g_1+g_2 \over 2 m_p^2}\,S_{ij}\nabla_i
\nabla_j\right)\,{e^{ipr} \over r}\,.
\eeq
By the analytical continuation of wave function (\ref{wf}) to the
point $p=i\kappa$, which corresponds to the bound state, we arrive
at the following expression for the deuteron wave function at
large distances:
\beq\label{dd}
\Psi_d=\,\sqrt{{\kappa \over 2\pi}}\left(1+{g_1+g_2 \over 2 \al_t
m_p^2}\,S_{ij}\nabla_i \nabla_j\right)\,{e^{-\kappa r} \over r}\,.
\eeq
The expectation value of the quadrupole moment tensor in this
state is
\beq\label{tqf}
Q_{ij}=\,{1 \over 4}\,\langle \, 3 r_i r_j - r^2 \delta_{ij} \,
\rangle =\,{g_1+g_2 \over \al_t m_p^2}\,S_{ij}.
\eeq
Now the deuteron quadrupole moment is
\beq\label{qf}
Q=Q_{zz}|_{S_z=1}=\,{g_1+g_2 \over 3 \al_t m_p^2}\,.
\eeq
Let us note that, since the principal contribution to $Q$ is given
by $r \sim 1/\kappa$, it does not make sense to introduce the
correction factors $(1-\kappa r_t)^{-1/2}$ and $(1-\kappa
r_t)^{-1}$ into expressions (\ref{dd}) ¨ (\ref{tqf}),
respectively.

The analysis~\cite{bkk,er} of the experimental data on the elastic
$dp$ scattering and on the stripping reaction results in
\beq
g_1+g_2=67\;\,\mbox{fm}.
\eeq
The corresponding value of the deuteron quadrupole moment is
\beq\label{q1}
Q_1=0.18\;\,\mbox{fm$^2$}.
\eeq
Somewhat larger values follow from the phase analysis~\cite{ar} of
the $np$ scattering:
\beq\label{q2}
Q_1=(0.20\; -\; 0.24)\;\,\mbox{fm$^2$}.
\eeq
At least the last value, (\ref{q2}), is not so far from the result
of the direct experimental measurement of the deuteron quadrupole
moment,
\beq\label{qe}
Q=0.286\;\,\mbox{fm$^2$}.
\eeq

Of course, the accuracy of the direct experimental result
(\ref{qe}) for the deuteron quadrupole moment is much higher then
the accuracy both of the approximations made for the derivation of
formula (\ref{qf}) and of the phase analysis of the mentioned
elastic processes and the stripping reaction. Therefore, the
relations obtained, besides their theoretical interest, may be
useful mainly for checks of phenomenological descriptions of these
reactions.

\section{Circular polarization of $\gamma$-quanta\\ in the reaction
$np \to d\gamma$ with polarized neutrons}

Just as the photodisintegration of the deuteron at the threshold
goes into the $^1S_0$ wave of continuous spectrum, the radiative
capture of thermal neutrons $np \to d\gamma$ proceeds from the
same $^1S_0$ state via an $M1$ transition. Naturally, in this
approximation, when the initial $^1S_0$ state is completely
spherically-symmetric, the polarization of an initial particle
cannot be transferred to a final particle. However, due to the $D$
wave admixture in the deuteron wave function and in the incoming
wave, the $M1$ transition proceeds also from the triplet initial
state. Besides, the corrections to the $M1$ operator responsible
for the nonadditivity of nucleon magnetic moments in the deuteron
also allow for a magnetic dipole transition from the triplet
initial state. And at last, due to the $D$ wave admixture an $E2$
transition becomes possible as well. All these effects are fairly
small, but their investigation gives us an information on some
subtle details of the $np$ interactions at low energies.

The circular polarization of the $\gamma$-quanta emitted in the
radiative capture of polarized thermal neutrons by unpolarized
protons $np \to d\gamma$ was for the first time measured
in~\cite{ves}. This result,
\[ P_{\gamma}=-(2.90 \pm 0.87)\times 10^{-3}, \]
was improved essentially in the next work of the same
group~\cite{baz}:
\beq\label{exp}
P_{\gamma}=-(1.5 \pm 0.3)\times 10^{-3}.
\eeq
Theoretically the problem has been considered in articles~[13-18].
In the present review we essentially follow works~\cite{da2,buk}.

We start with the magnetic moment operator
\[ {\bf M}=\,{e \over 2 m_p}\,({\bf l}_p + \mu_p \si_p +\mu_n
\si_n). \] The proton orbital angular momentum ${\bf l}_p$ is
obviously related to the total orbital angular momentum ${\bf L}$:
${\bf l}_p=1/2{\bf L}$. The linear combination of ${\bf L}$ and
$\si_{p,n}$ can be presented as follows:
\[
{1 \over 2}\,{\bf L} + \mu_p \si_p +\mu_n \si_n \]
\beq\label{dec}
 = {1 \over 2}\left({\bf L}+{1 \over 2}\,\si_p + {1 \over 2}\,\si_n\right) +{1
\over 2}\,(\mu_p - \mu_n)(\si_p - \si_n) +{1 \over 2}\left(\mu_p +
\mu_n-\,{1 \over 2}\right)(\si_p + \si_n)
\eeq
The first term in the second line is nothing but one half of the
total angular momentum $\vI$ which, being an integral of motion,
cannot cause any transition. The second term is responsible for
the common $M1$ transition from the $^1S_0$ state of the
continuous spectrum, i.e., for the process inverse to the deuteron
photodisintegration at the threshold. The effective operator of
this, principal $M1$ transition, or the coordinate matrix element
of the second term, is
\beq\label{m0}
\hat{\bf M}_0=\,{e \over 2 m_p}\,{1 \over 2}\,(\mu_p - \mu_n)
\sqrt{2\pi\kappa}\,{1-\kappa \al_s \over \kappa^2}\,(\si_p -
\si_n).
\eeq

As to the last term in the second line of (\ref{dec}), its matrix
element is distinct from zero first of all due to the $^3D_1$
admixture to the wave functions of the deuteron and the $^3S_1$
state of the continuous spectrum, $\psi_d$ and $\psi_{St}$,
respectively. For calculating this matrix element we use the
orthogonality condition for the complete radial wave functions of
the both states:
\beq\label{or}
\int_0^{\infty}dr r^2(R_{d0}R_{t0}+R_{d2}R_{t2})=0.
\eeq
Here we mark by the second subscripts, 0 and 2, the radial
functions of the $S$ and $D$ components of the initial and final
states. With the account for (\ref{or}) the discussed matrix
element is found easily:
\beq\label{me1}
{1 \over 2}\left(\mu_p + \mu_n-\,{1 \over 2}\right)\langle\,\si_p
+ \si_n\,\rangle =-\,{3 \over 2}\left(\mu_p + \mu_n-\,{1 \over
2}\right)\vI \int_0^{\infty}dr r^2 R_{d2}R_{t2}=0.
\eeq
The last radial integral is dominated by distances smaller than
the range of nuclear forces. It is natural to assume~\cite{da2}
that at these distances $R_{d2}$ and $R_{t2}$ differ in
normalization factors only, i.e., that
\[ R_{t2}=-\al_t\sqrt{{2\pi \over \kappa}}\,R_{d2}. \]
Then matrix element (\ref{me1}) reduces to
\beq
{3 \over 2}\left(\mu_p + \mu_n-\,{1 \over 2}\right)\al_t P_d \vI,
\eeq
where $P_d$ is the $D$ wave weight in the deuteron.

Let us recall now the known expression for the deuteron magnetic
moment $\mu_d$:
\beq\label{md}
\mu_d=\mu_p + \mu_n -\,{3 \over 2}\left(\mu_p + \mu_n-\,{1 \over
2}\right)P_d + \Delta \mu
\eeq
(numerically, $\mu_d=0.8574$). In line with the correction for the
$D$ wave admixture (the term with $P_d$), it includes the
contribution of relativistic corrections $\Delta \mu$. Naturally,
the matrix elements of corresponding operators are also dominated
by short distances. Let us assume again that at these distances
the wave function of the $^3S_1$ state of continuous spectrum
differs from the deuteron one in the overall factor only. Then the
coordinate matrix element of the corresponding operator of the
triplet-triplet transition is
\[ -\al_t\,\sqrt{{2\pi \over \kappa}}\,\Delta \mu \vI. \]
Arising in this way total effective operator of the $M1$
triplet-triplet transition is
\beq
\hat{\bf M}=\,{e \over 2 m_p}\,\left[\,{3 \over 2}\,\left(\mu_p +
\mu_n-\,{1 \over 2}\right)P_d - \Delta
\mu\,\right]\al_t\,\sqrt{{2\pi \over \kappa}}\,\vI,
\eeq
or, in virtue of (\ref{md}),
\beq
\hat{\bf M}=\,{e \over 2 m_p}\, (\mu_p +
\mu_n-\mu_d)\al_t\,\sqrt{{2\pi \over \kappa}}\,\vI.
\eeq

A standard calculation, taking into account expression (\ref{m0})
for the principal transition, leads to the following result for
the $M1$ contribution to the degree of circular polarization:
\beq\label{thm}
P_{\gamma}(M1)=-\,{\mu_p + \mu_n-\mu_d \over \mu_p - \mu_n}\,
{\kappa \al_t \over 1-\kappa \al_s} = - 0.92\times 10^{-3}.
\eeq

One more contribution to the degree of circular polarization of
photons is due to the quadrupole transition. The $E2$ operator
equals here
\[ V=-{1 \over 8}\, e r_m r_n \partial_m E_n  \]
(let us recall the relation $\br_p = \br/2$ between the proton
coordinate $\br_p$ and the argument $\br$ of the wave function).
Quite standard calculation with wave functions (\ref{wf}) and
(\ref{dd}) leads to the following result for the quadrupole
contribution to the circular polarization:
\beq\label{thq}
P_{\gamma}(E2)=-\,{2 \over 15}\,{\kappa^3 (g_1+g_2) \over m_p^2\,
(1-\kappa \al_s)\,(\mu_p-\mu_n)}\,=-\,{2 \over 5}\,{\al_t \kappa^3
Q \over (1-\kappa \al_s)\,(\mu_p-\mu_n)}\,=-0.25\times 10^{-3}.
\eeq

Our final result
\beq\label{th}
P_{\gamma}=P_{\gamma}(M1)+P_{\gamma}(E2)=-1.2\times 10^{-3}
\eeq
and the experimental result (\ref{exp}) are in a reasonable
agreement. There is also a good agreement between (\ref{th}) and
the final results of recent calculations~\cite{che,par}), though
for $P_{\gamma}(M1)$ and $P_{\gamma}(E2)$, taken separately, our
and their results differ considerably.

\section{P- and T-odd electromagnetic moments of deuteron}
\subsection{Generalities on anapole moments}
A notion of anapole moment (AM) was introduced by V.G. Vaks (who
was then a graduate student of A.B. Migdal) and Ya.B.
Zel'dovich~\cite{zev}. AM is a special electromagnetic
characteristic of a system where parity is not conserved.

The peculiarity of AM is in particular that the interaction of a
charged probe particle with an anapole moment is of a contact
nature (for a more detailed discussion see, for
instance,~\cite{kh}). Therefore, for example, the interaction of
the electron with the nucleon AM, being on the order of $\al G$
($G$ is the Fermi weak interaction constant), cannot be
distinguished in general case from other radiative corrections to
the weak electron-nucleon interaction. And in a gauge theory of
electroweak interactions only the total scattering amplitude,
i.e., the sum of all diagrams on the order of $\al G$, is
gauge-invariant, independent of the gauge choice for the Green's
functions of heavy vector bosons. Thus, generally speaking, the AM
of an elementary particle or a nucleus is not gauge-invariant
notion, and therefore has no direct physical meaning. However,
there are special cases where the anapole moment has a real
physical meaning. In heavy nuclei the AM is enhanced $\sim
A^{2/3}$~\cite{fks} ($A$ is the atomic number), as distinct from
common radiative corrections\footnote{It means, by the way, that
there is an intrinsic limit for the relative accuracy, $\sim
A^{2/3}$, with which the AM of a heavy nucleus can be defined at
all. In the case of $^{133}$Cs, which is of the experimental
interest, this limiting accuracy is about 4\%.}. In fact, the
nuclear AM of $^{133}$Cs was discovered and measured with good
accuracy in an atomic experiment~\cite{wi}. The result of this
experiment is in a reasonable quantitative agreement with the
theoretical predictions, starting with~\cite{fks,fk}.

There is one more object, the deuteron, whose anapole moment could
make sense for a sufficiently large P odd $\pi$NN
constant~\cite{fk}. The problem of the deuteron AM was considered
phenomenologically in~[1,23-25]. Recently the deuteron AM induced
by the P-odd pion exchange was calculated in~\cite{kk1} (see
also~\cite{ss}).

\subsection{Nucleon anapole moment}
It is convenient to start the discussion with the nucleon AM in
the limit $m_{\pi} \to 0$. It was calculated in 1980 by A.I.
Vainshtein and the author. The result is the same for the proton
and neutron:
\beq\label{nam}
\va_p\,=\,\va_n\,=\,-\,{e g \bar{g} \over 12 m_p m_{\pi}}\,
\left(1-\,{6 \over \pi}\,{m_{\pi} \over m_p}\ln{m_p \over
m_{\pi}}\right) \,\mbox{\boldmath $\sigma$};
\eeq
we assume $e>0$. Being the only contribution to the nucleon AM,
which is singular in $m_{\pi}$, the result (\ref{nam}) is
gauge-invariant. In this respect, it has a physical meaning.

Unfortunately, despite the singularity in $m_{\pi}$, the
corresponding contribution to the electron-nucleon scattering
amplitude is small numerically as compared to other radiative
corrections to the weak scattering amplitude. Indeed, the
radiative corrections to the effective constants $C_{2p,n}$ of the
proton and neutron axial neutral-current operators $G / \sqrt
2\,C_{2p,n}\mbox{\boldmath $\sigma$}_{p,n}$ are~\cite{ms}
\beq\label{cpn}
C^r_{2p}\,=\,-0.032\pm 0.030\,, \quad  C^r_{2n}\,=\,0.018\pm
0.030\,.
\eeq
In the same units $G/\sqrt 2$, the effective axial constants
induced by the electromagnetic interaction with the proton and
neutron anapole moments (\ref{nam}), is
\[ C^a_{p,n}\,=\,-\,\al a_N\,(|e|G/\sqrt 2)^{-1}\,=
\,0.07\times 10^5 \bar{g}. \] At the ``best value'' $\bar{g}=
3.3\times 10^{-7}$ (strongly supported by the experimental result
for the $^{133}$Cs anapole moment) we obtain
\beq\label{cna}
C^a_{p,n}\,=\, 0.002.
\eeq
With this value being much less than both central points and error
bars in (\ref{cpn}), the notion of the nucleon AM practically has
no physical meaning. This is why the result (\ref{nam}) was never
published by the authors. It is quoted in book~\cite{kh} (without
the logarithmic term) just as a theoretical curiosity. This result
was obtained also in~\cite{muh}, the logarithmic term in the
nucleon AM is discussed in~\cite{hhm}.

Let us note, that as pointed out in~\cite{kas}, a P-odd $\pi\pi$NN
interaction also generates a contribution $\sim \ln m_{\pi}$ to
the nucleon AM. Only purely theoretical estimates are known for
the constants of this P-odd $\pi\pi$NN interaction. According to
the estimates considered by the authors of~\cite{kas} as
relatively reliable, the contribution of the P-odd $\pi\pi$NN
interaction to the nucleon AM is about an order of magnitude
smaller than (\ref{nam}).

However, the situation with the deuteron AM is quite different.
Not only the proton and neutron AMs add up here. The isoscalar
part of the radiative corrections is much smaller than the
individual contributions $C^r_{2p}$ and $C^r_{2n}$, and is
calculated with much better accuracy~\cite{ms}:
\beq\label{crd}
C^r_{2d} = C^r_{2p} + C^r_{2n} = 0.014\pm 0.0030\,.
\eeq
Moreover, there are new large contributions $\sim m_{\pi}^{-1}$
to the deuteron AM, induced by the P-odd $\pi$-meson exchange. We
go over now to their consideration.

\subsection{P-odd $\pi$-meson exchange}
The Lagrangians of the strong $\pi$NN interaction and of the weak
P-odd one, $L_s$ and $L_w$, respectively, are well-known:
\begin{equation}\label{s}
L_s\,=\,g\,[\,\sqrt{2}\,(\overline{p}i\gamma_{5}n \,\pi^+
+\overline{n}i\gamma_{5}p\, \pi^-)\,+(\,\overline{p}i\gamma_{5}p
-\overline{n}i\gamma_{5}n)\,\pi^0];
\end{equation}
\begin{equation}\label{w}
L_w\,=\,\bar{g}\,\sqrt{2}\,i\,(\,\overline{p}n \,\pi^+
-\overline{n}p \,\pi^-).
\end{equation}
Our convention for $\gamma_5$ is
\beq\label{g5}
\gamma_5 = \left(\begin{array}{rr}0  & -I \\
 -I &  0 \end{array}\right);
\eeq
the relation between our P-odd $\pi$NN constant $\bar{g}$ and the
commonly used one $h^{(1)}_{\pi NN}$ is
$\bar{g}\sqrt{2}=h^{(1)}_{\pi NN}$. Our sign convention for the
coupling constants is standard: $g=13.45$, and $\bar{g} > 0$ for
the range of values discussed in~\cite{ddh}.

The resulting effective nonrelativistic Hamiltonian of the P-odd
nucleon-nucleon interaction in the deuteron due to the pion
exchange is in the momentum representation
\beq\label{q}
V(\bq)\,=\,{2 g \bar{g} \over m_p}\,{(\vI \bq) \over m_{\pi}^2 +
\bq^2} (N_1^{\dagger}\tau_{1-}N_1)\,(N_2^{\dagger}\tau_{2+}N_2).
\eeq
Here $\bq =\p_1'-\p_1= -(\p_2'-\p_2)
 =\p_n'-\p_p= -(\p_p'-\p_n).$
This P-odd interaction conserves the total spin
\[ \vI = {1 \over 2}\, (\mbox{\boldmath $\sigma$}_p
+\mbox{\boldmath $\sigma$}_n) \] (and does not conserve the
isotopic spin). Thus in our problem it mixes the $^3P_1$ state of
the continuous spectrum and the deuteron ground state $^3S_1$. Let
us note that the P-odd interaction (\ref{q}) which interchanges
the proton and neutron, when applied to the initial state
$a^{\dagger}_p(\br_1)a^{\dagger}_n(\br_2)|0\rangle$ transforms it
into $a^{\dagger}_n(\br_1)a^{\dagger}_p(\br_2)|0\rangle
=-a^{\dagger}_p(\br_2)a^{\dagger}_n(\br_1)|0\rangle$. On the other
hand, the coordinate wave function of the admixed $^3P_1$ state is
proportional to the relative coordinate $\br$, which we define as
$\br_p-\br_n$. Therefore, it also changes sign under the
permutation $p \leftrightarrow n$. Thus, for the deuteron the
P-odd potential can be written in the coordinate representation as
a simple function of $\br=\br_p-\br_n$ without any indication of
the isotopic variables:
\beq\label{r}
V(\br)\,=\,{g \bar{g} \over 2\pi m_p}\,(-i\vI \cdot
\mbox{\boldmath $\nabla$}) \,{\exp{(-m_{\pi}r)} \over r}\,.
\eeq

The corresponding imaginary mixing matrix elements are related as
follows:
\[
\langle ^3P_1 | V | ^3S_1 \rangle = - \langle ^3S_1 | V | ^3P_1
\rangle,
\]
the overall sign in (\ref{r}) corresponds to $\langle ^3P_1 | V |
^3S_1 \rangle$.

The weak interaction (\ref{r}) generates a contact current ${\bf
j}^c$. To obtain an explicit expression for it, we have to
consider the P-odd interaction $V$ in the presence of the
electromagnetic field. Its including modifies the proton
momentum: $\p \to \p - e \vA$, which results in the shift $\bq \to
\bq + e\vA$ in interaction (\ref{q}). Then in the momentum
representation the contact current is
\[ \bj^c(\bq)\,=\,-\,{\partial V(\bq) \over \partial \vA}\,=\,
-\,{\partial \over \partial \vA}\;{2g\bar{g} \over m_p}\; {\vI
(\bq + e \vA) \over m_{\pi}^2 + (\bq+ e \vA)^2}\, \]
\beq\label{jc}
=\, -\,{2 e g\bar{g} \over m_p}\;\left\{{\vI \over m_{\pi}^2 +
\bq^2}\, -\,{2 \bq (\vI \bq) \over (m_{\pi}^2 + \bq^2)^2}\right\}.
\eeq
In the last expression we have neglected the dependence on $\vA$.
In the coordinate representation the contact current is
\beq\label{cc}
\bj^c(\br)\,=\,{e g\bar{g} \over 2\pi m_p}\,\br(\vI \Na)\,
{e^{-m_{\pi}r} \over r}\,.
\eeq

\subsection{Calculation of the deuteron anapole moment}
Let us discuss at first a general structure of the deuteron AM
generated by a P-odd $np$ interaction, which conserves the total
spin, assuming only that the deuteron is a pure $^3S_1$ state
bound by a spherically symmetric potential. We follow here
essentially the arguments applied in~\cite{fk} (see also
book~\cite{kh}) to the problem of a single proton in a spherically
symmetric potential. In this case of a single proton the AM
operator is~\cite{fk}
\beq\label{gap}
\hat{\va}={\pi e \over m_p}\left\{\mu_p\, \br \times \si -\,{i
\over 3}\,[{\bf l}^2, \br]\right\} +\,{2\pi \over 3}\,\br \times
[\br \times \bj^c\,].
\eeq
In the case of the deuteron this formula generalizes to
\beq\label{gad}
\hat{\va}_d={\pi e \over 2 m_p}\left\{\br \times (\mu_p\si_p -
\mu_n\si_n) -\,{i \over 6}\,[{\bf l}^2, \br]\right\} +\,{\pi \over
6}\,\br \times [\br \times \bj^c\,].
\eeq
Both operators (\ref{gap}) and (\ref{gad}) are orthogonal to $\br$
(neither of them commutes with $\br$, so the orthogonality means
here that $\hat{\va}\br+\br\hat{\va}=0$). Therefore the contact
current (\ref{cc}), generated by the P-odd pion exchange and
directed along $\br$, does not contribute to the nuclear AM.

Let us present the wave function of the deuteron $^3S_1$ state as
$\psi_0(r)\chi_t$. If the weak interaction conserves the total
spin $\vI$, the P-odd admixture $^3P_1$ can be written as
$\delta\psi_1(\br)=i(\vI\br/r)\psi_1(r)\chi_t$ (both radial wave
functions, $\psi_0(r)$ and $\psi_1(r)$, are spherically
symmetric). Simple calculations demonstrate that the expectation
value of operator (\ref{gad}) in the state with wave function
$[\psi_0(r)+i(\vI\br/r)\psi_1(r)]\chi_t$ is (in the absence of a
contact current contribution)
\beq\label{amd}
\va_d={\pi e \over 3 m_p}\left(\mu_p  - \mu_n  - {1 \over
3}\,\right) \vI \int d \br \psi_0(r)  r  \psi_1(r).
\eeq
So, under the assumptions made, the deuteron AM should depend on
the universal combination $\mu_p - \mu_n - 1 / 3$.

The real calculations we perform in the zero-range approximation
where $\psi_0(r)=\psi_d(r)$ (see (\ref{d})), and the wave
functions of $P$ states are free. Moreover, when using the
stationary perturbation theory, we can choose plane waves as
intermediate states we are summing over, since perturbation
$V(\br)$ (see (\ref{r})) will select by itself $P$ states from
plane waves. The correction to the wave function is
\beq\label{ps1}
\delta\psi_1(\br)\,=\,\int {d \bk \over (2\pi)^3}\; {e^{i \bk \br}
\over -\ep - k^2/m_p}\; \int d \br'\,e^{- i \bk \br'} V(\br')\,
\psi_0(r').
\eeq
Rather lengthy calculation leads to the following expression for
the matrix element of the radius-vector:
\beq\label{rv}
\int d \br \psi_d(r) \br \delta\psi_1(\br)\,=\,-i\,{e g\bar{g}\vI
\over 6 \pi m_{\pi}} \,{1+\xi \over (1+2\xi)^2}\,,
\eeq
where $\xi=\kappa/m_{\pi}=0.32$. With this matrix element and
operator (\ref{gad}) one obtains easily the following expression
for the deuteron AM:
\beq\label{fa}
\va_d^{(0)}\,=\,-\,{e g\bar{g} \over 6 m_p m_{\pi}} \,{1+\xi \over
(1+2\xi)^2}\,\left(\mu_p-\mu_n- \,{1 \over 3}\right)\vI\,,
\eeq
in accordance with the general formula (\ref{amd}).

In fact, the range $1/m_{\pi}$ of the P-odd interaction (\ref{r})
is quite comparable to the range of the usual nuclear forces.
Therefore, it is rather dangerous to use here blindly the
na\"{\i}ve wave function (\ref{d}). Still, numerical calculations
with a model wave function, which has more realistic properties,
lead to the result which differs from (\ref{fa}) no more than by
20\%. As to other sources of P-violation, different from the pion
exchange, it can be demonstrated that their contribution to the
deuteron AM does not exceed 5\% of (\ref{fa}) (at the ``best
value'' $\bar{g}$).

It looks reasonable to combine the potential contribution
(\ref{fa}) with the sum of the proton and neutron anapole moments
(\ref{nam}). In this way we arrive at the final result for the
deuteron AM:
\beq\label{ffa}
\va_d =\,-\,{e g\bar{g} \over 6 m_p m_{\pi}}\,
[\,0.49(\mu_p-\mu_n-1/3)\,+\,0.46\,]\,\vI  =\,-\,2.60\,{e g\bar{g}
\over 6 m_p m_{\pi}}\,\vI.
\eeq
This result includes all contributions to the P-odd amplitude of
$ed$-scattering, which are singular in $m_{\pi}$, and thus is
gauge-invariant, physically sensible.

Now let us compare the contribution of the found AM (\ref{ffa})
with the P-odd $ed$-scattering amplitude due to usual radiative
corrections, which are nonsingular in $m_{\pi}$. At the ``best
value'' $\bar{g}= 3.3\times 10^{-7}$ and the estimate of 20\% for
the accuracy we obtain
\beq\label{ca}
C^a_{2d}\,=\,0.014 \pm 0.003.
\eeq
This number is quite comparable with the contribution (\ref{crd})
of usual radiative correction. Taken together, they constitute
\beq\label{ctd}
C_{2d}=C^r_{2d}+C^a_{2d} = 0.028\pm 0.005\,.
\eeq
One more contribution to $C_{2d}$ is due to the admixture of
strange quarks in nucleons~\cite{ef}. The magnitude of this
contribution is both extremely interesting and highly uncertain.

To measure experimentally the constant $C_{2d}$ is very difficult.
However, with such a good accuracy of the theoretical prediction
(\ref{ctd}), this experiment would be a source of valuable
information on the P-odd $\pi$NN constant and on the strange
quark admixture in nucleons.

\subsection{P-odd, T-odd electromagnetic moments of the deuteron}
P-odd, T-odd multipoles of the deuteron, electric dipole and
magnetic quadrupole, were previously considered phenomenologically
in~\cite{sfk}. Then in~\cite{kk1} these multipoles have been
calculated in the same approach as the anapole moment.

Three independent P-odd, T-odd effective $\pi$NN Lagrangians are
conveniently classified by their isotopic properties:
\beq\label{0}
\Delta T =0. \quad L_0\,=\,g_0\,[\,\sqrt{2}\,(\overline{p}n
\,\pi^+ +\overline{n}p\, \pi^-)\,+(\,\overline{p}p
-\overline{n}n)\,\pi^0];
\end{equation}
\begin{equation}\label{1}
|\Delta T| =1. \quad
L_1\,=\,g_1\,\bar{N}N\,\pi^0\,=\,g_1\,(\,\overline{p}p \,+
\,\overline{n}n \,)\,\pi^0);\quad \quad \quad \quad
\end{equation}
\[ |\Delta T| = 2. \quad L_2\,=\,g_2\,(
\,\bar{N}\mbox{\boldmath $\tau$}N\,\mbox{\boldmath $\pi$}\,-
\,3\bar{N}\tau^3 N\,\pi^0\,) \quad \quad \quad \quad \quad \quad
\]
\beq\label{2}
\quad \quad \quad \quad \quad \quad \quad \quad \quad \;\;\;
=\,g_2\,[\,\sqrt{2}\,(\overline{p}n
\,\pi^+\,\overline{n}p\,\pi^-)\, -\,2\,(\,\overline{p}p
\,-\,\overline{n}n \,)\,\pi^0)\,].
\end{equation}
Since the possible values of the isotopic spin for two nucleons is
$T=0,\,1$ only, the last interaction, with $|\Delta T| = 2$, is
not operative in our approach.

The effective P-odd, T-odd $np$ interaction is derived in the same
way as (\ref{r}). In the momentum representation it looks as
follows:
\beq\label{ww}
W(\br)\,=\,{g \over 8\pi m_p}\, [\,(\,3g_0 -
g_1\,)\,\mbox{\boldmath $\sigma$}_p\,-\, (\,3g_0 +
g_1\,)\,\mbox{\boldmath $\sigma$}_n\,] \,\mbox{\boldmath
$\nabla$}\,{e^{-m_{\pi}r} \over r}\,.
\eeq

The calculation of the deuteron electric dipole moment ${\bf
d}_d$, i.e., of the $e\br_p=e\br/2$ expectation value, is
performed analogously to that of the AM. The result is
\beq
{\bf d}\,=\,-\,{egg_1 \over 12\pi m_{\pi}}\,{1+\xi \over
(1+2\xi)^2}\,\vI.
\eeq

The magnetic quadrupole moment (MQM) operator is expressed through
the current density $\bj$ as follows (see, e.g.,~\cite{kh,kl}):
\beq\label{m}
M_{mn}\,=\,(r_m\ep_{nrs}+ r_n\ep_{mrs})r_r j_s.
\eeq
This expression transforms to
\beq\label{mm}
M_{mn}\,=\,{e \over 2m}\left\{\,3\mu\left[\,r_m\sigma_n +
r_n\sigma_m -\,{2 \over 3}\,(\mbox{\boldmath
$\sigma$}\br)\right]\,+ 2q (r_m l_n + r_n l_m)\right\}.
\eeq
Here $\mu$ is the total magnetic moment of the particle, $q$ is
its charge in the units of $e$. The magnetic quadrupole moment is
the expectation value ${\cal M}$ of the operator $M_{zz}$ in the
state with the maximum spin projection $I_z=I$.

In our case, due to the spherical symmetry of the deuteron
nonperturbed wave function, the orbital contribution to $M_{mn}$
vanishes. The contact current here also is directed along $\br$
and therefore does not contribute to MQM. Thus the deuteron MQM
arises due to the spin term in (\ref{mm}). It equals
\beq
{\cal M}\,=\,-\,{e g \over 12\pi m_p m_{\pi}}\,{1+\xi \over
(1+2\xi)^2}\, [\,(\,3g_0+g_1\,)\,\mu_p +
(\,3g_0-g_1\,)\,\mu_n\,)\,].
\eeq

Let us note in conclusion that the possibility to search for the
deuteron dipole moment at a storage ring with polarized nuclei is
discussed at present by experimentalists rather seriously.

\begin{center}***\end{center}
I am grateful to A.V. Blinov, A.P. Burichenko, L.A. Kondratyuk,
and R.V. Korkin, the review is based on the results obtained
together with them. The work was supported by the Grant for
Leading Scientific Schools No. 00 15 96811; by the Ministry of
Education through Grant No. 3N-224-98; by the Federal Program
Integration-1998 through Project No. 274.

\end{document}